\documentstyle[aaspptwo]{article}

\def \n {\noindent}
\begin{document}

\centerline {\bf Cumulant Correlators in 2D and 3D Scale Free Simulations }

\bigskip
\bigskip
\centerline{Dipak Munshi$^1$ and Adrian L. Melott$^2$}
\bigskip
\centerline{$^1$ Astronomy Unit, Queen Mary and Westfield College, London E1
4NS, United Kingdom. }
\centerline{$^2$ Department of Physics and Astronomy, University of Kansas,
Lawrence, Kansas 66045, U.S.A. }
\smallskip
\centerline {Email: D.Munshi@qmw.ac.uk, melott@kusmos.phsx.ukans.edu}

\bigskip
\topskip 1.0cm
\centerline{\bf Abstract}
\vskip .5cm
Shape dependence of higher order correlations
introduces complication in direct determination of these quantities.
For this reason theoretical and observational progress has been
restricted in calculating
one point distribution functions and  related moments.
Methods based on factorial moments of two point count probability distribution
(cumulant correlators) were recently shown to be efficient in subtracting
discreteness effects and
 extracting useful information from galaxy catalogs. We use
these cumulant correlators $Q_{NM}$ to study
clustering in scale free simulations both in two and three dimensions. Using
this method in the highly nonlinear regime we were able to
separate hierarchal amplitudes $Q_3, R_a ~{\rm and}~ R_b$ associated with different
tree graphs
contributing to third and fourth order correlation functions. They were found
to increase with  power on large scales.  Results based on
factorial moments of one point cumulants show very good agreement with
that on cumulant correlators. Comparison of simulation results with
perturbation theory and extended perturbation theory were found to be in
reasonable agreement. Comparisons
were also made against  predictions from hierarchal models for higher order
correlations. We argue that finite volume corrections are very important for
computation of cumulant correlators.
\bigskip

\n
{\it Subject Headings:large scale structure of the universe - galaxies:
statistics - methods: data analysis - methods: simulation}

\bigskip
\centerline{\bf 1. Introduction}
\bigskip
\n
There is growing evidence which suggest that the large scale structure in the
universe was formed by the gravitational amplification of small
inhomogeneities.  Statistical characterization of  clustering is  important
for understanding the dynamics of gravitational clustering.  Correlation
functions and count in cell statistics are among the oldest and most
widely used statistics to  quantify clustering. Evaluation of higher order
correlation
functions with full shape dependence is difficult given limited size of
present day galaxy catalogs and numerical simulations (Fry \& Peebles 1978,
Peebles 1980). Much progress has therefore been made in estimation of
volume averages of these quantities
 (Gaztanaga 1992, Bouchet et al. 1993, Gaztanaga \& Frieman 1994, Colombi et al. 1996a, Szapudi et al. 1996), which can be directly related to moments of one
point count probability distribution function. Analytical methods based on tree
level perturbation 
theory and loop corrections
were used to evaluate these quantities in the quasi-linear regime (Peebles 1980,
Juszkiewicz et al. 1993, Bernardeau 1992, Bernardeau 1994, Munshi et al. 1994,
Colombi et al. 1992, Scoccimarro \& Frieman 1996a-b,). In the highly nonlinear regime, although no full theory
exists, scaling models are often used to predict their behaviour (Peebles 1980,
Balian \& Schaeffer 1989). Computational methods were also developed based on factorial
moments to subtract Poisson noise from discrete data. Errors associated with
these one point cumulants and other related quantities such as void probability
function have also been estimated and correction procedure have been
developed (Colombi et al. 1995, Szapudi \& Colombi 1996, Munshi et al. 1997).

Although one point quantities carry useful information about dynamics and
background
geometry averaging associated with such quantities causes a significant loss of
information. Alternative methods use correlation of pairs of cells
(Szapudi et al. 1992, Meiksin et al. 1992, Szapudi et al. 1995). Cumulant
correlators provide a natural generalization of one point cumulants.
The hierarchal ansatz in the strong clustering regime and perturbation theory
in the weak clustering regime have definite predictions for these quantities.
Using cumulant correlators it is possible to separate amplitudes associated
with different tree topologies up to fifth order in the correlation hierarchy.
Szapudi \& Szalay (1997) have analysed APM data using cumulant correlators. We
use these statistics to study scale--free simulations in two and three
dimensions (2D and 3D).

In the next section we outline the theoretical framework necessary for using
cumulant correlators, predictions from perturbation theory, hierarchal ansatz
and extended perturbation theory. In section $\S$3 we describe
 our numerical simulations and techniques to evaluate
cumulant correlators from simulation data. We discuss implications of our
results in section $\S$4 and compare it with existing results from galaxy
catalogs.

\bigskip
\vfill
\centerline{\bf 2. Theoretical Predictions }
\bigskip
\n
\bigskip
We define factorial moment correlators of a pair of cells of volume $l^3$
separated by a distance $r$ as
\begin{equation}
w_{kl}(l,r) = {{\langle (N_1)_k(N_2)_l\rangle - \langle (N_1)_k \rangle
\langle(N_2)_l\rangle} \over  \langle N\rangle ^{ k + l}}; ~~~k\ne 0, l\ne 0,
\end{equation}
\noindent
and the normalized one point factorial moment

\begin{equation}
w_{k0}(l) = {\langle  (N_1)_k\rangle \over \langle N_1\rangle^k },
\end{equation}

\noindent
where we have used the notation $(N)_k = N(N-1)\dots(N-k + 1)$, and the angular
braces $\langle \dots \rangle$  denote average over all possible positions of
the cells.

It is possible to relate $w_{k0}$ and $w_{kl}$ to normalized one point
cumulants $S_N =  \langle \delta^N \rangle / \langle \delta^2 \rangle^{(N-1)}$
and cumulant correlators $C_{NM} = \langle \delta^N({\bf x}_1) \delta^M({\bf
x}_2)
\rangle / \langle \delta({ \bf x}_1) \delta ({\bf x}_2)\rangle \langle \delta^2
\rangle^{(N+M-2)}$. However we find it to easier to work with $Q_N =
S_N/N^{(N-2)}$ and $Q_{NM} = C_{NM}/N^{(N-1)}M^{(M-1)}$. While central moments
can also be used to computed these quantities, factorial moments are known
to be more suitable for subtracting Poisson shot noise.

One can introduce the generating function for the factorial moments in terms
of the cumulants $Q_N$.

\begin{equation}
W(x) = \exp \sum_{N=1}^{\infty} \Gamma_N x^N Q_N
\end{equation}

\noindent
where we have defined
$\Gamma_N = {N^{N-2} {\xi_s}^{N-1}/ N!}$ and  ${\xi_s}$ is
the volume average of $\xi_2$ over volume of the cell $l^3$.

The generating function can be linked with the one point void probability
distribution function. In general $Q_N$ parameters show scale dependence,
increasing from
quasi-linear phase to highly nonlinear phase. Hierarchal form of correlation
functions demand $Q_N$ to be independent of scale in the nonlinear regime.

We can also construct generating function of factorial moments
$W(x,y) = \sum_0^{\infty}W_{mn}{x^M y^N/m!n!}$ which can be related to
generating function of cumulant correlators $Q(x,y)$
\begin{equation}
Q(x,y) = \xi_l \sum_{ M = 1, N = 1}^{\infty} x^M y^N Q_{NM} \Gamma_M \Gamma_M
NM,
\end{equation}
\n
by the following equation
\begin{equation}
W(x,y) = W(x)W(y)(\exp Q(x,y) - 1).
\label{taylor}
\end{equation}



\n
Expanding equation (\ref{taylor}) one can compute cumulant correlators $Q_{NM}$
from factorial moments $W_{NM}$.

\begin{eqnarray}
&Q_{12}2\Gamma_1\Gamma_2\xi_r = w_{12}/2 - \xi_r& \nonumber \\
&Q_{13}3\Gamma_1\Gamma_3\xi_r  = w_{13}/6 - w_{12}/2 - w_{20}/2 +
\xi_r&\nonumber \\
&Q_{22}\Gamma_2^24\xi_r = w_{22}/4 - w_{12} + \xi_r - \xi_r^2/2&
\label{taylor1}
\end{eqnarray}

As it is clear from equations (\ref{taylor1}), cumulant correlators are
determined by two different length scales, length scale corresponding
to ${\xi}_s$ and scale corresponding to ${\xi}_r$. We use the
terms quasi-linear and nonlinear depending upon values of $\xi_s$.

\centerline{\bf 2.1 Nonlinear Regime}

\n
The extreme nonlinear stage of gravitational clustering is believed to be
well described by a power law solution for correlation function $\xi_2 =
r^{-\gamma}$, where $\gamma$ can be expressed as a function of initial power
spectrum (Davis \& Peebles 1977, Balian \& Schaeffer 1989.  All
higher order correlations exhibit a self similar scaling in strong clustering
regime

\begin{equation}
\xi_N( \lambda {\bf r}_1, \dots  \lambda {\bf r}_N ) = \lambda^{-\gamma(N-1)}
\xi_N( {\bf r}_1, \dots {\bf r}_N ).
\label{hierarchal}
\end{equation}

More explicit functional forms for $N$-point correlations are constructed
by summing over products of $N-1$ two-point correlation functions corresponding
to different topologies, each of
which representing a  tree graph spanning N-points with amplitudes
$T_{n,\alpha}$

\begin{equation}
\xi_N( {\bf r}_1, \dots {\bf r}_N ) = \sum_{\alpha, \rm N-trees} T_{N,\alpha}
\sum_{\rm labelings}
\prod_{\rm edges}^{(N-1)} \xi_{(r_i, r_j)}.
\end{equation}

Tree graphs spanning three points can have only one topology. The amplitude
associated with this graph $Q_3$ contributes through three different
configurations of these points, so that cumulant correlators to have following
form:

\begin{equation}
\langle \delta({\bf x}_1)^2 \delta({\bf x}_2) \rangle = 2Q_{12}\xi_s \xi_r =
Q_3(2\xi_s\xi_r+\xi_r^2 ).
\end{equation}

Higher order cumulants get contributions from different types of tree diagrams.
At fourth order trees connecting four points have two different topologies,
known as the snake (with its amplitude denoted by $R_a$) and the star topology
(corresponding amplitude denoted conventionally by $R_b$). Using this
notation
we can write

\begin{eqnarray}
\langle \delta_1({\bf x}_1)^3 \delta_2({\bf x}_2)\rangle = 9Q_{13}\xi_r\xi_s^2
=
6\xi_r \xi_s^2 R_a + 3 \xi_r \xi_s^2 R_b + 6\xi_r^2 \xi_s R_a + \xi_r^3R_b\\
\langle \delta({\bf x}_1)^2 \delta({\bf x}_2)^2 \rangle = 4Q_{22}\xi_r\xi_s^2 =
4\xi_r \xi_s^2 R_a + 4 \xi_r^2 \xi_s R_b + 4\xi_r^2 \xi_s R_a + 4\xi_r^3R_b.
\label{topology}
\end{eqnarray}

These simultaneous equations are linear in $R_a$ and $R_b$ and can be solved
once $Q_{22}$ and $Q_{13}$ are determined. When $\xi_r<<1$  which will
be the case when two cells are far away, one can define linear cumulant
correlators (linear in $\xi_r$)  by considering only terms linear in $\xi_r$.
The linear solution
to equation (\ref{topology}) are $R_a = Q_{22}$ and $R_b = 3Q_{13} - 2Q_{22}$.
In such a linear accuracy one can show that

\begin{equation}
Q_{NM} \approx Q_{N+M}
\label{other}
\end{equation}

Although amplitudes of tree terms with different topologies can be estimated
by cumulant correlators, the
situation becomes more complicated at higher order as number of degenerate
correlators become less than number of topologies, making the system
indeterministic. Other interesting questions regarding scaling of generating
functionss $Q(x,y)$ and related question about the nature of $Q_{NM}$ in the
highly nonlinear regime can be addressed when bigger N-body simulations become
available.

\bigskip
\centerline{\bf 2.2 Quasi-linear Regime}
In quasi-linear regime when $\xi_s$ is smaller than unity it is possible to
expand $\delta = \delta^{(1)} + \delta^{(2)} + \delta^{(3)} \dots $ where
perturbation
expansion is valid as long as the series is convergent. Using perturbative
calculation of two point quantities it was possible for Bernardeau (1996) to
express $C_{pq}$ at lowest order. Perturbative terms contributing to
the expansion of $\langle \delta^p({\bf x}_1) \delta^q({\bf x}_2)\rangle$
can be written as

\begin{equation}
\langle\delta^p({\bf x}_1) \delta^q({\bf x}_2)\rangle \approx
\sum_{\rm decompositions} \langle \prod_{i=1}^p \delta^{(p_i)}({\bf x}_1)
\prod_{j=1}^q\delta^{(q_i)}({\bf x}_2) \rangle
\end{equation}

\n
where sum is taken over all possible decompositions and $p_i$ and $q_i$
satisfies

\begin{equation}
\sum_{i=1}^p p_i + \sum_{j=1}^q q_j = 2(p+q)-2.
\end{equation}

\n
In lowest order in perturbation theory it is possible to write (Bernardeau
1996),

\begin{equation}
\langle \delta^p({\bf x}_1) \delta^q({\bf x}_2) \rangle_c = C_{pq} \langle
\delta^2(x)
\rangle^{p+q-2}\langle \delta(x_1) \delta(x_2) \rangle = C_{pq}\xi_s^{p+q-2}
\xi_r.
\end{equation}

\n
Using method of generating function it was shown that  $C_{pq}$
can be decomposed into following relation (Bernardeau, 1996),

\begin{equation}
C_{pq} = C_{p1}C_{q1}.
\end{equation}

\n
In 3D the lowest order $C_{pq}$ can be expressed in terms of spectral index
$n$,

\begin{eqnarray}
&C_{21} = {68 / 21} - {1 / 3} ( n+3)&\\
&C_{31} = {11710 / 441} - {61 / 7}(n+3) + {2 / 3}(n+3)^2.&
\end{eqnarray}

\n
It should be noted that such factorization is possible only in case of
tree-level diagrams also calculations were done using large separation
limit. Given the whole hierarchy of $C_{NM}$ it is possible to
compute bias associated with gravitational clustering in the quasi-linear
regime (Bernardeau 1996).

\bigskip
\centerline{\bf 2.3 Connecting different regimes: Extended Perturbation Theory}

Tree level perturbation theory can predict $S_N$
parameters for top-hat smoothing, it was used (as out lined above) to compute
cumulant correlators $Q_{N,M}$ to arbitrary order in large separation
limit. The expressions derived for cumulants depends on index of initial power
spectrum. However
it was later realized that the same quasi-linear expressions can  describe
evolution of $S_N$ parameters from quasi-linear regime to the nonlinear
regime
by allowing spectral index $n$ to vary with scale (Colombi et al. 1996b). It is
interesting
that whole hierarchy of $S_N$ can be described by such a variation of
effective spectral index $n_{eff}$ (unfortunately no straight forward relation
exists between  true nonlinear power spectra and $n_{eff}$). An extension of
perturbation theory to the non--perturbative regime was considered by Szapudi \&
Szalay (1997) for
cumulant correlators $Q_{N,M}$. Tree level perturbation theory as mentioned
before predicts $C_{pq} = C_{p1}C_{q1}$.
Using same logic as extended perturbation theory one can expect such a relation
to hold even for small cell sizes at large distances although separately each
of these term may vary significantly.

\begin{figure}
\protect\centerline{
\epsfysize = 7.55truein
\epsfbox[20 146 587 714]
{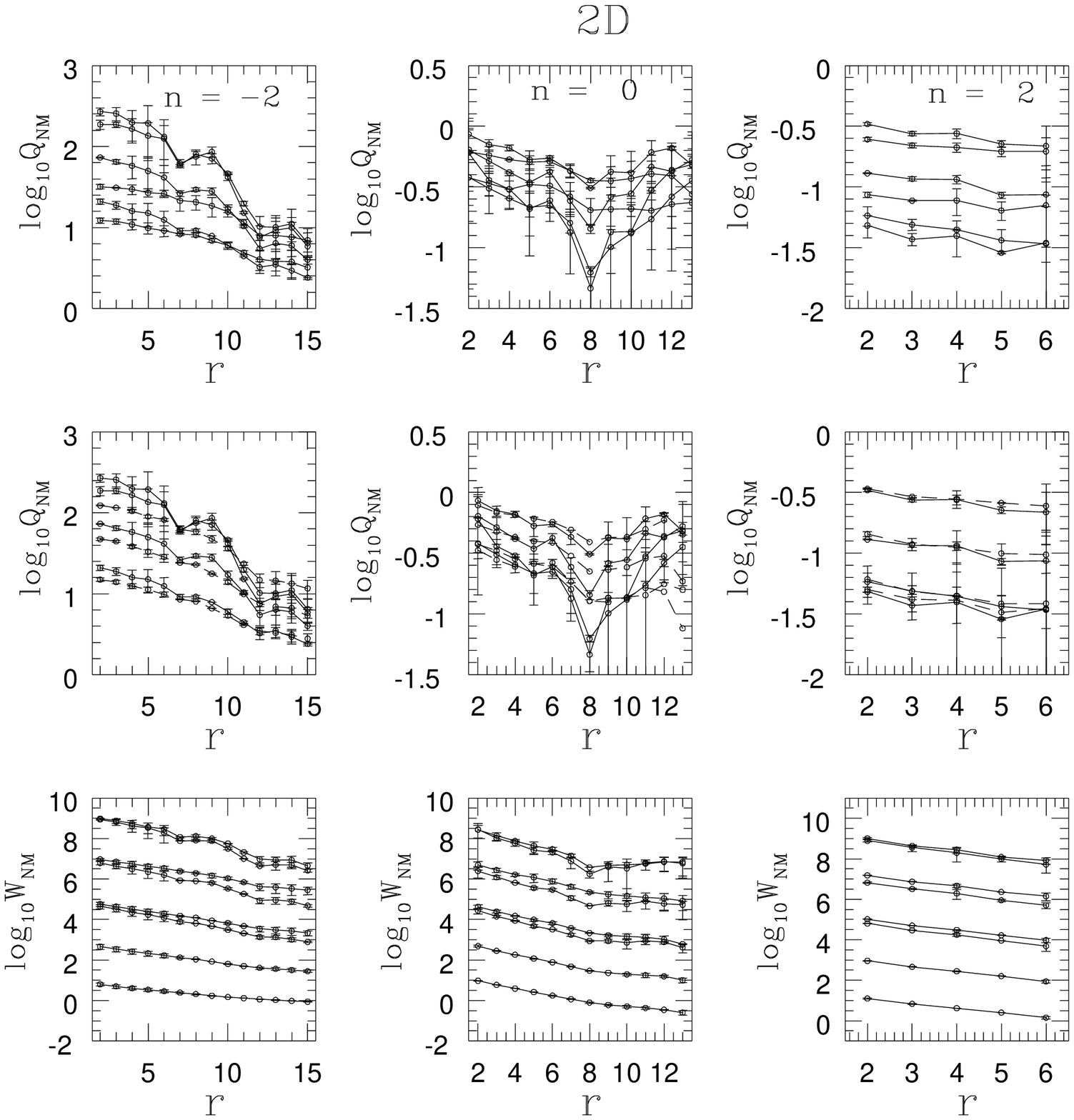}
}
\caption{ Lower panel displays factorial moment correlators $w_{kl}$ as
a function of separation $r$ in 2D in units of grid scale. Degenerate
parallel lines correspond to increasing values of $N+M$ from $2$ to $6$.
Left panels display results for scale free spectra with power law index
$n=-2$, middle panel for $n=0$ and right panel $n=2$ in two dimension.
Middle and upper panels plot cumulant correlators of different orders.
Dashed curves in middle panels
represent $Q_{N1}Q_{M1}$ for different values of $N$ and $M$,
closest solid curves represent $Q_{NM}$ for same values of $N$ and $M$.
According to predictions from extended perturbation theory they should
overlap $Q_{NM} = Q_{N1}Q_{M1}$. Top most panels compare predictions
of hierarchal ansatz $Q_{NM} = Q_{N+M}$. Adjacent curves correspond to
different $N$ and $M$ with same $N+M$. Error bars were calculated
by finding scatter in results with different realization of same initial
power spectra.}
\end{figure}

\begin{figure}
\protect\centerline{
\epsfysize = 7.55truein
\epsfbox[20 146 587 714]
{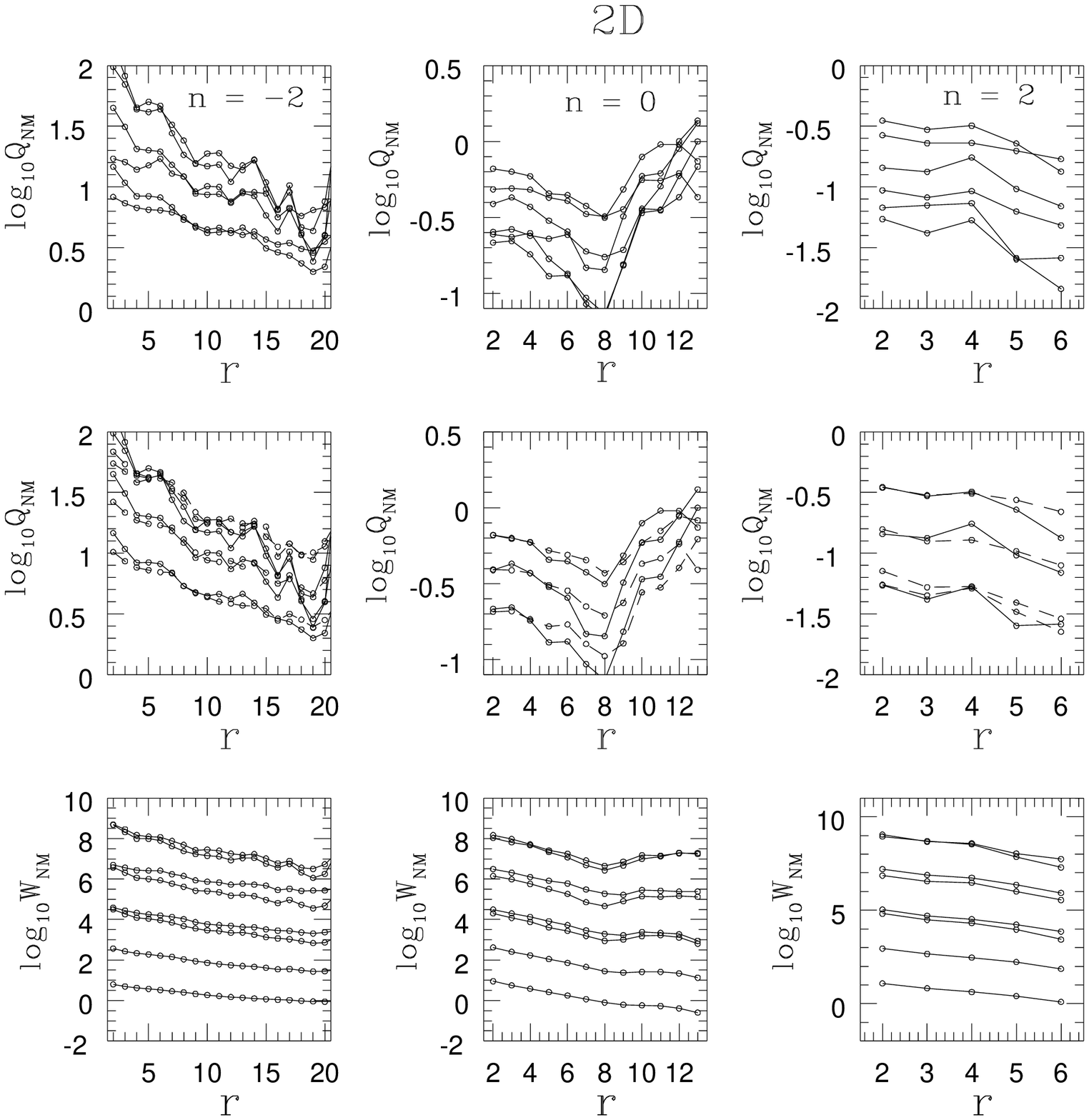}
}
\caption{ Same as Figure - 1 but results from only one realization are
plotted to show level of agreement with theoretical predictions in individual
realizations.  }
\end{figure}

\begin{figure}
\protect\centerline{
\epsfysize = 2.10truein
\epsfbox[23 530 584 721]
{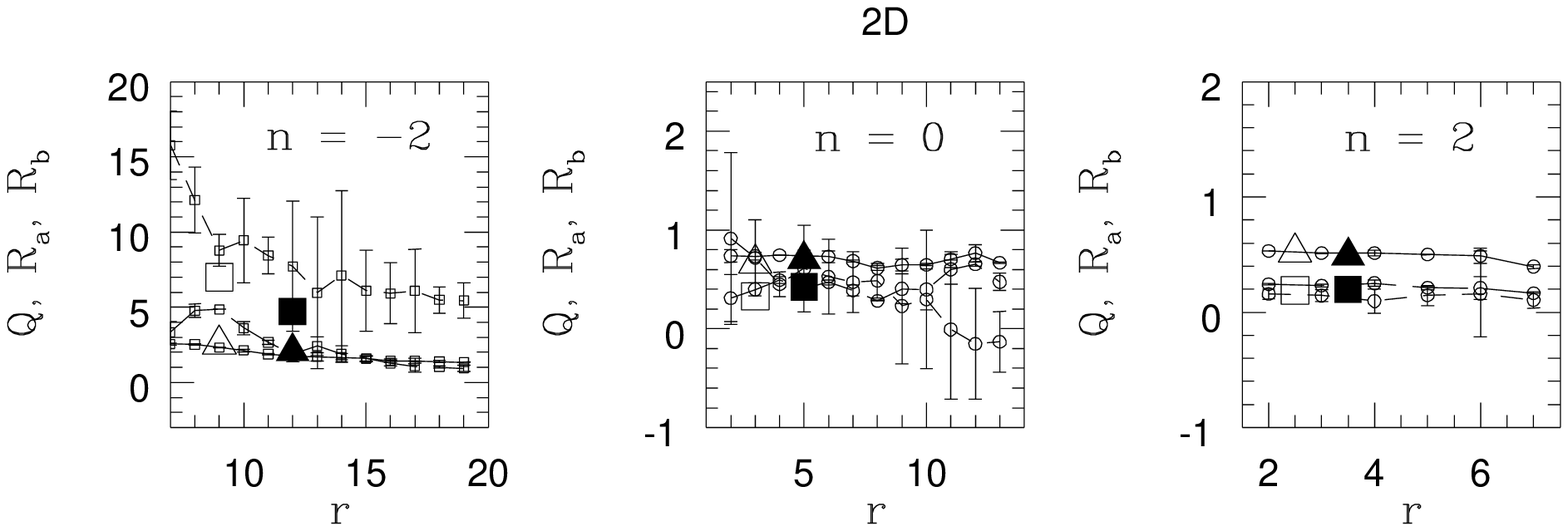}
}
\caption{ Lower order  hierarchal amplitudes $Q_3$, $R_a$, $R_b$ calculated
from the fully nonlinear cumulative correlators in 2D are
plotted against separation $r$ measured in grid units. Solid curve
represent estimates of $Q_3$ lower and upper dashed lines represent
amplitudes of fourth order snake and star graphs $R_a$ and $R_b$
respectively. Open triangles and square correspond measurement of $Q_3$
and $Q_4$ respectively from factorial moments of counts in cell
statistics after finite volume corrections were taken into account (from Munshi
et al. 1997). Filled
triangles and squares are measurements of same quantities using cumulant
correlators without finite volume correction.}
\end{figure}

\bigskip
\vfill
\centerline{\bf 3. Simulations and Data Analysis }
\bigskip
\n
The simulations used here are numerical models for the gravitational
clustering of  collisionless  particles in an expanding background. We study
evolution of initial
Gaussian perturbations in $\Omega = 1$ universe. All the 2D simulations are
done with a particle-mesh (PM) code with $512^2$ particles with an equal number
of grid points and in 3D $128^3$ particles with $128^3$ grid points
The code has at least twice the dynamical
resolution of any other PM code with which it has been compared.
The 2D simulations are described in detail in Beacom et. al (1991), with
a video of the evolution in Kauffmann and Melott (1992).  The 3D simulations
are described in Melott and Shandarin (1993).  Both sets, which consist
of multiple realizations (different random seeds) for a variety of power
spectra, have been widely used for comparative studies of various statistical
methods and dynamical approximations. 

In this paper we analyze a subset of the simulations
with featureless power-law initial spectra of the general form,
\begin{eqnarray}
P(k)&\propto&\ k^n\ {\rm for}\ k \le k_c,\\
&=&~ 0\ {\rm for}\  k > k_c.
\end{eqnarray}
We have analyzed power-law models with $n= 2, 0, -2$ in 2D and
$n= 1, 0, -1, -2$ in 3D with a cutoff in each case at the Nyquist wave
number: $k_c = 256\,k_f$ for 2D and $k_c = 64\,k_f$ for 3D where
$k_f={2 \pi/ L_{\rm box}}$ is the fundamental mode associated with
the box size.

We choose $\sigma(k_{\rm NL})$, the epoch when the scale $2\pi/k_{\rm NL}$
is going nonlinear as a measure of time.

\begin{figure*}
\protect\centerline{
\epsfysize = 3.3truein
\epsfbox[20 146 587 714]
{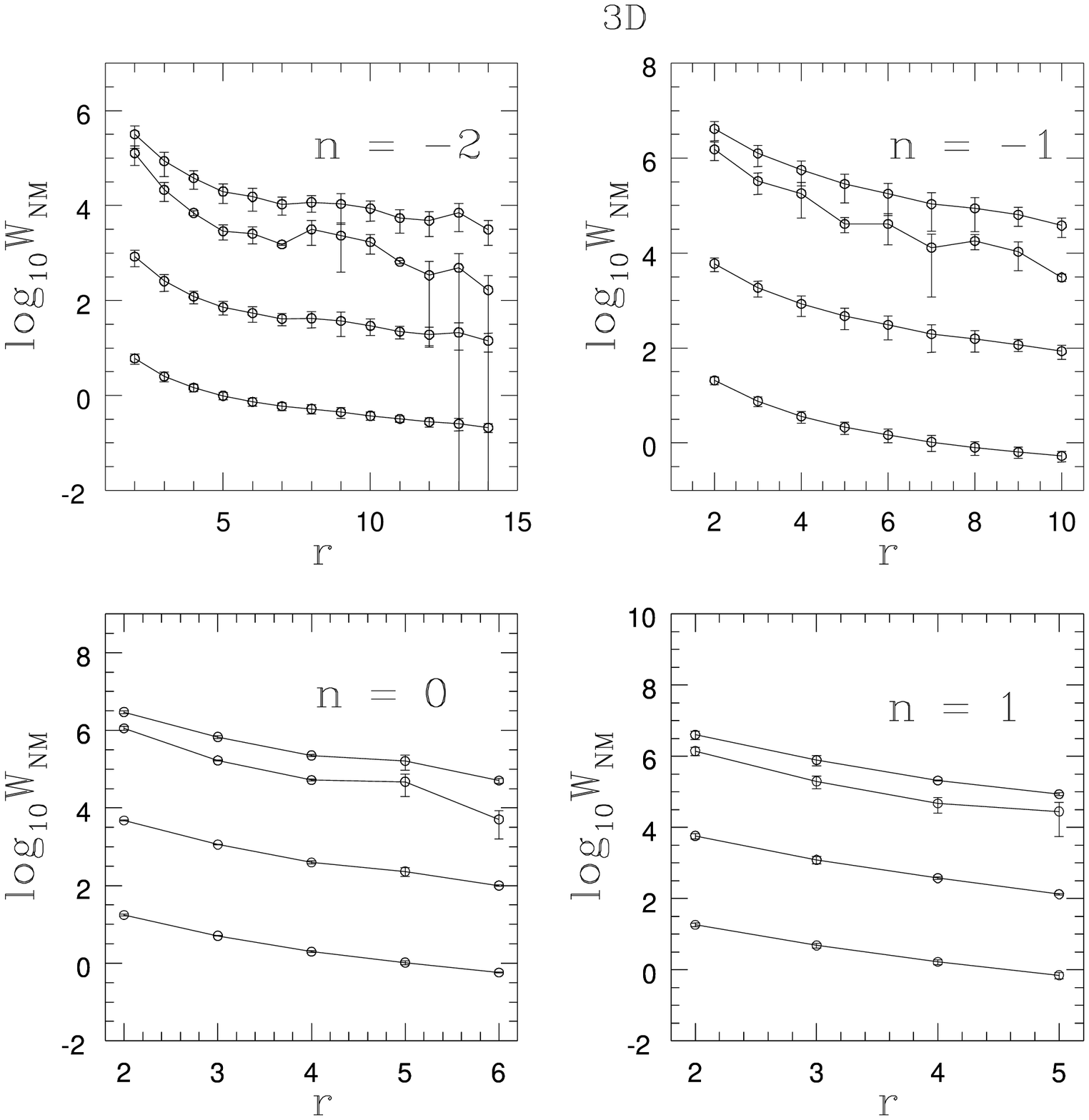}
}
\caption{ Each panel displays factorial moment correlators $w_{kl}$ as
a function of separation $r$ in unit of grid scale for different initial
scale--free power law spectra in 3D. Scatter in each plot
is computed from scatter in four different realizations of same power
spectra.}
\end{figure*}

\begin{figure*}
\protect\centerline{
\epsfysize = 3.3truein
\epsfbox[20 146 587 714]
{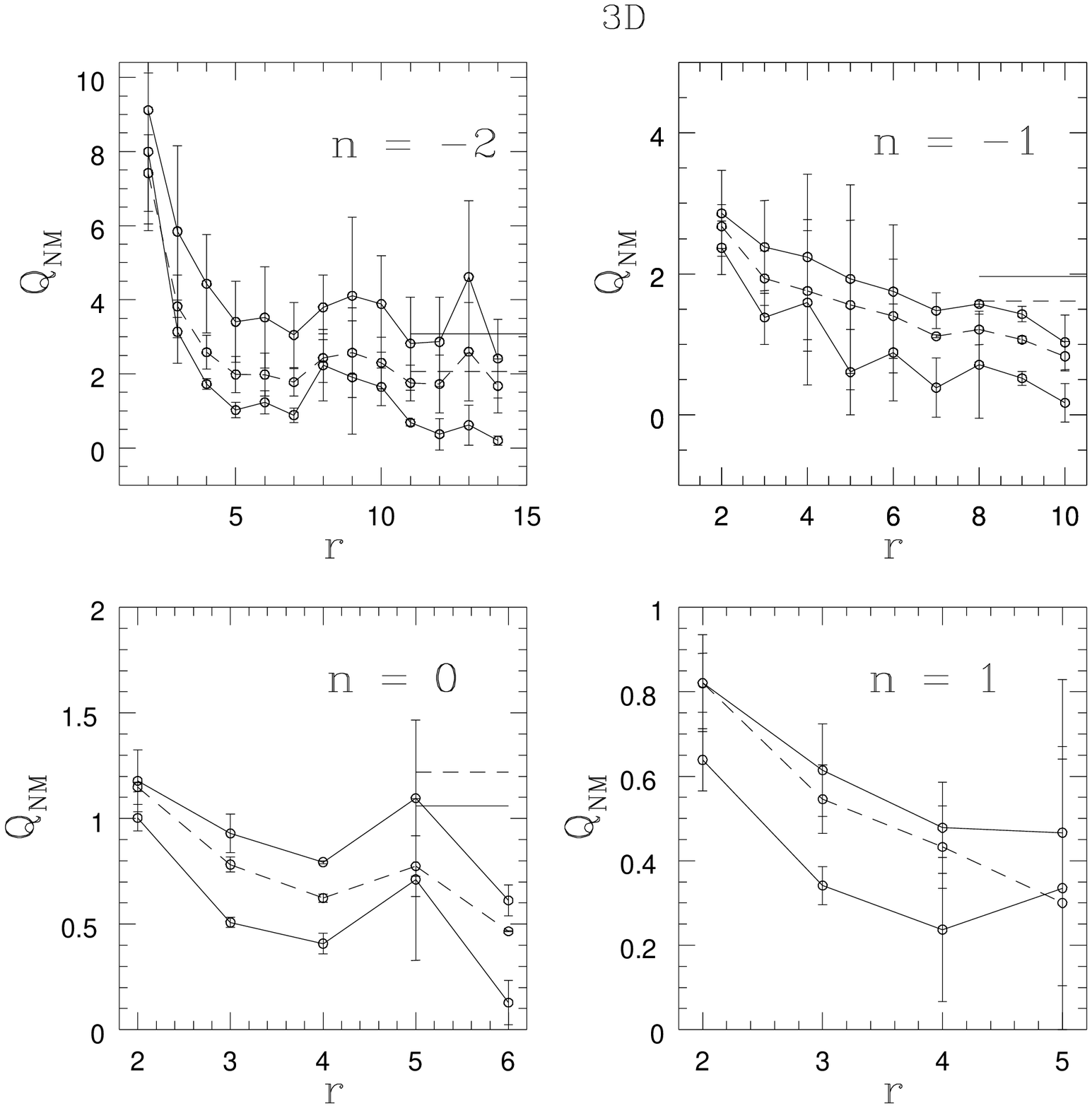}
}
\caption{ Nonlinear cumulant correlators is plotted against separation $r$
in 3D as measured in units of grid scale. Uppermost curve correspond to
$Q_{31}$, lowest curve to $Q_{22}$ and the dashed one to $Q_{21}^2$. Solid
and dashed straight lines  correspond to predictions for $Q_{31},~\rm and~
Q_{21}^2$ from perturbations theory at large separation.}
\end{figure*}

\begin{figure*}
\protect\centerline{
\epsfysize = 3.3truein
\epsfbox[20 146 587 714]
{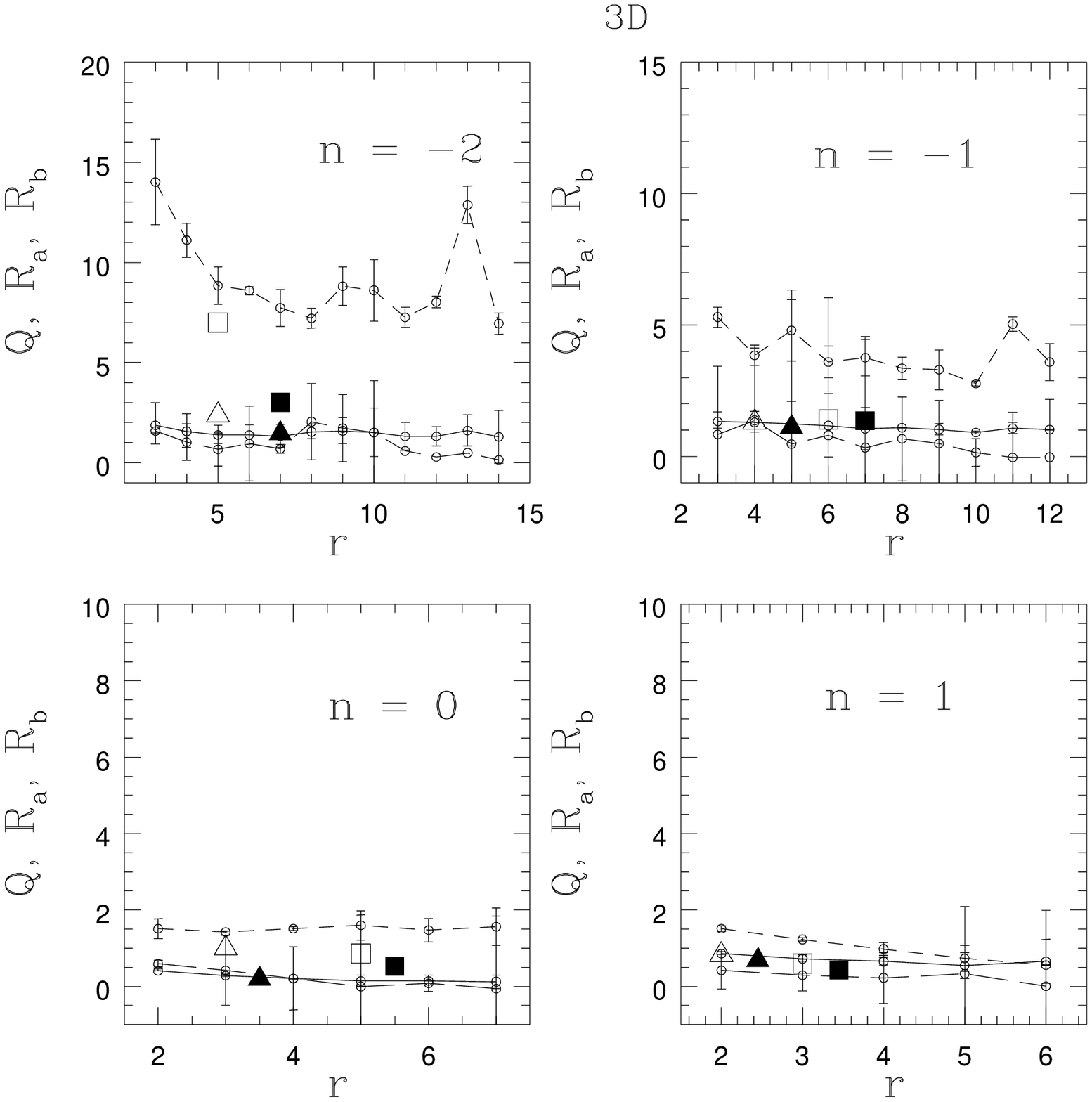}
}
\caption{ Hierarchal amplitudes $Q_3$, $R_a$, $R_b$ calculated
from the  nonlinear cumulative correlators in 3D are
plotted against separation $r$ measured in grid units. Solid curve
represent estimates of $Q_3$ lower and upper dashed lines represent
amplitudes of fourth order snake and star graphs $R_a$ and $R_b$
respectively. Open triangles and square correspond to measurement of $Q_3$
and $Q_4$ respectively from factorial moments of counts in cell
statistics after finite volume corrections were taken into account (from Munshi
et al. 1997). Filled
triangles and squares are measurements of same quantities using cumulant
correlators without finite volume correction.}
\end{figure*}

\begin{equation}
 \sigma ( k_{\rm NL}) = \left({{\int_{k_f}^{k_{\rm Ny}} P(k)\,k\,\d k } \over
{\int_{k_f}^{k_{\rm NL}} P(k)\,k\,\d k }}\right)^{ 1 \over 2}
\end{equation}

The first scale to go nonlinear is the one corresponding to the Nyquist
wave number. This happens, by definition, when the variance
$\sigma$ is unity.
Of course as $\sigma$
increases successive larger scales enter in the nonlinear regime.
The simulations were stopped at $\lambda_{\rm NL}=2l_{\rm grid},
\ 4l_{\rm grid},\ 8l_{\rm grid}, ....,\ L_{\rm box}/2$.
In our study in 2D we took the epoch when  $L_{\rm box}/16$ for analysis
of $n= -2$ spectrum and for $n = 0 ~\rm~ and 2$ we studied the epoch
when $L_{\rm box}/8$ is going nonlinear. In 3D we had less dynamic range so
choose the epoch when  $L_{\rm box}/4$ for analysis
of $n= -2$ spectrum and for $n = -1, 0, ~\rm~ and 1$ we studied the epoch
when $L_{\rm box}/2$ is going nonlinear. The separation $r$ that we study is
much less compared to the length scale going nonlinear so it is unlikely
that our results will be affected much by boundary conditions.

The growth rate of various modes in the linear regime were studied by
Melott et al. (1988) for
this PM code. The results 
at $ \lambda = 3l_{\rm grid} $ are equivalent
to the ones obtained by
a typical PM code at $\lambda = 8l_{\rm grid} $, due to
the staggered mesh scheme. So we expect that our code performs well at the
wavelength associated with four cells and since the collapse of
$4\,l_{\rm grid}$-size
perturbations will give rise to condensations of diameter $ 2\,l_{\rm grid} $
or less, the smallest cell size
that can be safely resolved is $ 2\,l_{\rm grid}$.

Computations of count probability distribution function (CPDF) $P_l(n)$ and
two-point count probability distribution $P_{l,r}(n,m)$ were done by
laying down a grid of mesh spacing $l$ and counting occupation number in
each cell to compute the probability of finding $n$ objects in cell size $l$
and also the joint probability of finding $n$ and $m$ objects in two cells
separated by a distance $r$. Statistics were improved by perturbing the
grid in each orthogonal direction and keeping the mesh undistorted while
repeating the counting process. Both in two and three dimensions we
considered only cells of size $l_{grid}$. While this particular choice
of cell size is open to question, we will show that our results match
remarkably well with our earlier studies done using bigger cell sizes.
With this cell size we could reach probabilities as few times $10^{-6}$
in 2D and 3D. Computations of factorial moments
and factorial moment correlators were done using computed values of $P_l(n)$
and $P_{l,r}(n,m)$. Computed values of $\bar \xi_2$ for cells used
were found to be $8.98, 27.87, 60.39$ in 2D for spectral index $-2, 0,~\rm and~2$
respectively. In 3D these values were found to be $25.46, 88, 125.5, ~\rm and~
171$ respectively.

Different spurious effects such as shot noise and finite volume corrections
are important while computing count probability distribution functions.
For small cell sizes Poisson noise starts dominating, whereas larger
cell sizes are dominated by finite volume corrections. We find $w_{kl}$
parameters to be dominated by shot noise with increasing separation of
cells, this effects starts to be severe with higher order cumulants.
This effect starts dominating even for small separation for  spectra with less
power on larger scales. Using large cells produces effect similar to smoothing
the density field and hence reducing the effective correlation
length scale between them. These two dominant effects reduce the range of
separation
which can be probed for studying cumulant correlators. Finite volume
corrections
are more difficult to quantify. Methods based on scaling of count in cell
statistics were shown to be effective in correcting finite volume effects
(Munshi et al. 1997),
similar arguments can be used to develop corrections for cumulant correlators.
The validity of such a method depends on correctness of the scaling model,
which we  test in this paper. Developments of methods to correct
$Q_{NM}$ for finite volume effect are left for future work.

Computed $w_{kl}$ for 2D simulations are plotted in Figure - 1 as a function of
cell separation $r$. Corresponding results for 3D are presented in Figure - 5.
Parallel degenerate lines correspond to different values of $N+M$
increasing from bottom to top in each panel. Middle and top panels
show variation of $Q_{NM}$ with $r$. Middle panels
test the validity of extended perturbation theory. Dashed lines correspond to
$Q_{N1}Q_{M1}$ while neighboring solid lines represent $Q_{NM}$
for same values of $N$ and $M$. Error bars are computed by
estimating the scatter in results from four different realization of
each spectra. It is interesting to note that
while $Q_{NM}$ shows an increasing trend with increasing $N+M$ for spectra
with large scale power $n=-2$ in 2D, the trend is reversed for $n=0 ~\rm and~
2$. The results presented in 2D are for $N+M = 4, 5 ~\rm and~  6$ respectively.
In 3D meaningful computation of $Q_{N,M}$ were possible only for
$N+M < 4$. We find that computed values of $Q_{NM}$ are more stable
towards fluctuation at large separation if one of the indices of cumulant
correlators is larger than the other (e.g. $w_{31}$ is more stable at larger
separation as compared to $w_{22}$). Close associations of dashed curves with
solid curves proves the validity of extended perturbation theory. It is also
easy to notice that agreement is better in case of 2D compared to 3D.
Larger simulation size might be a possible explanation for such an effect.
Given that we restrict our result to highly nonlinear
regime, it is intersting to note that computed values of $C_{NM}$ are
not completely different from predictions of perturbation theory at 
large separation. In Figure - 5, dashed and solid straight lines are predicted
values of $Q_{21}^2$ and $Q_{31}$ from equation (17) and equation (18).
Top panels in Figure - 1 and Figure - 5 tests the validity of equation
(\ref{other}) in 2D and 3D respectively. Nearest neighbor curves correspond
to different values of $N$ and $M$ which produces same $N+M$. Agreement in this
case is
similar to that of extended perturbation theory.

Using cumulant correlators it is possible to compute amplitudes of
different tree topologies contributing to four and five point correlation
functions. We have computed amplitudes $Q$, $R_a$ and $R_b$. Figure - 3
shows that these amplitudes are almost constant in 2D in the highly nonlinear
regime. In 3D Figure - 6
shows large fluctuations in computed values of $R_b$ although in general
they are fairly constant within the limited range of nonlinearity studied by
us. Open triangles and
squares denote measured values of $Q_3$ and $Q_4$ using factorial moments
of CPDF. On the other hand solid triangles and squares denotes measured
values of $Q_3$ and $Q_4$ from cumulant correlators. $Q_4$ was calculated by
using the relation $16Q_4 = 12R_a + 4R_b$, from computed average values of
$R_a$ and $R_b$. The remarkable agreement of both the methods increases
confidence in results based on cumulant correlators. It is also to be noted
that whereas
computation of one point cumulants were done by taking volume corrections
into account, such effects were neglected in the computation of cumulant
correlators.
Our results show that $Q_4$ is much closer to $R_b$ which might indicate that
star topologies in general dominate over snake topologies.
However, more systematic studies are necessary which will incorporate
finite volume effects and have more dynamic range than this study.

\bigskip
\centerline {\bf 6. Discussion}
\bigskip

\n
Measurements of cumulant correlators and lower order one point cumulants
have already been done for Lick catalog, SDES and the APM survey. Analyzing
cumulant
correlators in the APM survey, Szapudi \& Szalay (1997) demonstrated the
validity of the hierarchal ansatz
to unprecedented accuracy. They found $Q_3 =1.15$, $R_a = 5.3$ and $R_b =
1.15$. These results were also shown to be in good agreement with
computation of one point cumulants based on factorial moments $Q_3 = 1.7$ and
$Q_4 = 4.17$. Measurements
form SDES give $Q_3 = 1.16$ and $Q_4 = 2.96$. It was suggested that slightly
low value of $Q_4$ form SDES was lack of nonlinear corrections in their
computations. Other measurements from APM produces $Q_3 = 1.6$ Szapudi et al.
(1996) and $Q_3 = 1.7$  by Gaztanaga (1994). The values of fourth order
cumulants found by these authors were $Q_4 = 3.2$ and $Q_4 = 3.7$
respectively.

Use of methods based on cumulant correlators by Szapudi and Szalay has
produced lower values for $R_b$ compared to $R_a$.  In our
simulations both in 2D and 3D we find that  relative position of
these two amplitudes depend on initial power spectra. In  all
cases $R_b$ was found to be either equal to or greater than $R_a$, although
the separation was found to decrease with increasing $n$. For power
spectra such as
$n=1$ in 3D and $n=2$ in 2D which have more power at smaller scales these two
quantities are found to almost coincide with each other. Although this is in
disagreement with findings
of Szapudi and Szalay (1997), our results seems to be closer to that of by Fry
\& Peebles (1978) who found $R_a = 2.5 \pm 0.6 ~\rm and~ R_b = 4.3 \pm 1.2$
from their analysis of Lick catalog.
At any rate, they are analyzing data in projection, and we a simulation
with full information.  We do not know to what extent projection and the
possible inclusion of non--gravitational physics may influence the results.

There have been several attempts to solve BBGKY  equations
in the highly nonlinear regime, although no general solution exists so far.
Attempts were made to solve these equations by assuming specific separable
 hierarchal forms for higher order correlations in phase space. Fry (1982)
derived $Q_{N,\alpha} = Q_N = (4Q/N)^{N-2} { N /(N-2)}$, in particular this
model
predicts $R_a = R_b = Q_4= 2Q^2/3$. Hamilton (1988) redoing the analysis found
only
snake topologies to contribute in higher order correlations while all non-snake
contributions vanish identically which gives $Q_N = (Q/N)^{N-2}{N! /2}$.
For lower order amplitudes this
predicts $R_a = Q^2, R_b= 0 ~\rm and~Q_4 = 3Q^2/4$. The ansatz used by
Bernardeau
and Schaeffer (1992) predicts $R_a = Q_3^2$. While all these efforts to solve
BBGKY
 equations provide useful insight
into complications arising due to the nonlinear nature of the
problem, it is clear from our study that cumulant correlators are a
very useful tool in comparing such predictions with computer simulations.
The general agreement of 2D and 3D results reinforces the many other results
which have shown that 2D may be a useful tool to extend the dynamic range
for exploration of statistics in cosmology.

\n
{\bf Acknowledgements:}
D.M. acknowledges support from PPARC under the QMW Astronomy Rolling
Grant GR/K94133. A.L.M. wishes to acknowledge the National Center for
Supercomputing
Applications for support to perform the ensemble of simulations, and
the financial support of the NSF EPSCoR program.

\bigskip

\bigskip

\end{document}